\documentclass[aps, onecolumn, superscriptaddress,showpacs]{revtex4}
\usepackage{graphicx,subfigure,epsfig}
\usepackage{amsmath,amssymb}
\input{epsf}
\input{epsfx}
\usepackage{epsfig}

\def\ket#1{\mathinner{|{#1}\rangle}}
\def\braket#1{\mathinner{\langle{#1}\rangle}}

\newcommand{\fr}[2]{\frac{{#1}}{{#2}}}

\newcommand{\unit}[1]{\hspace{0pt} {#1}}

\begin{document}
\title{Generation and transfer of single photons on a photonic crystal chip}

\date \today

\author{Dirk Englund}
\affiliation{Ginzton Laboratory, Stanford University, Stanford CA 94305}
\author{Andrei Faraon}
\affiliation{Ginzton Laboratory, Stanford University, Stanford CA 94305}
\author{Bingyang Zhang}
\affiliation{Ginzton Laboratory, Stanford University, Stanford CA 94305}
\author{Yoshihisa Yamamoto}
\affiliation{Ginzton Laboratory, Stanford University, Stanford CA 94305}
\author{Jelena Vu\v{c}kovi\'{c}}
\affiliation{Ginzton Laboratory, Stanford University, Stanford CA 94305}


\begin{abstract}
We present a basic building block of a quantum network consisting of a quantum dot coupled to a source cavity, which in turn is coupled to a target cavity via a waveguide.  The single photon emission from the high-$Q/V$ source cavity is characterized by a twelve-fold spontaneous emission (SE) rate enhancement that results in a SE coupling efficiency $\beta\sim0.98$ into the source cavity mode.  Single photons are efficiently transferred into the target cavity through the waveguide, with a source/target field intensity ratio of 0.12 (up to 0.49 observed in other structures without coupled quantum dots).  This system shows great promise as a building block of future on-chip quantum information processing systems.
\end{abstract}
\maketitle

\begin{figure}
     \epsfig{file=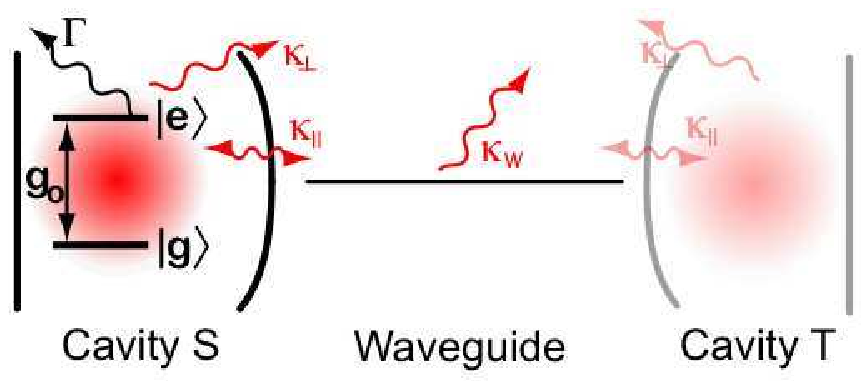, width=4in}
    \caption{Basic network consisting of two cavities and one cavity-coupled QD.  The QD 2-level system is coupled to a cavity (with coupling $g_0$) and decays with SE rate $\Gamma$.  The cavity, in turn, is coupled to a waveguide and leaky modes at field coupling rates $\kappa_{\|}$ and $\kappa_{\perp}$, respectively.  The waveguide field decay rate is $\kappa_W$.     }
    \label{fig:Fig1}
\end{figure}

Recent years have witnessed dramatic practical and theoretical advancements towards creating the basic components of quantum information processing (QIP) devices.  One essential element is a source of single indistinguishable photons, which is required in quantum teleportation\cite{Bouwmeester97nature}, linear-optics quantum computation\cite{KLM01}, and several schemes for quantum cryptography \cite{NC00}.  Sources have been demonstrated from a variety of systems\cite{SVG2005Europhysics} including semiconductor quantum dots (QDs)\cite{Michler03book}, whose efficiency and indistinguishability can be dramatically improved by placing it inside a microcavity\cite{Santori02}.  A second major component is a quantum channel for efficiently transferring information between spatially separated nodes of a quantum network\cite{CZKM1997PRL}.  This network would combine the ease of storing and manipulating quantum information in quantum dots\cite{PhysRevLett.83.4204}, atoms or ions\cite{PhysRevLett.75.4714,Wineland2004Nature}, with the advantages of transferring information between nodes via photons, using coherent interfaces\cite{FattalThesis2005,Sham2005PRL,Monroe2004Nature}.  Here we demonstrate a basic building block of such a quantum network by the generation and transfer of single photons on a photonic crystal (PC) chip.   A cavity-coupled QD single photon source is connected through a $25$\unit{$\mu$m} channel to an otherwise identical target cavity so that different cavities may be interrogated and manipulated independently (Fig. \ref{fig:Fig1}).  This system provides a source of single photons with a high degree of indistinguishability (mean wavepacket overlap of $\sim 67$\%), 12-fold spontaneous emission (SE) rate enhancement, SE coupling factor $\beta \sim 0.98$ into the cavity mode, and high-efficiency coupling into a waveguide.  These photons are transferred into the target cavity with a target/source field intensity ratio of 0.12, showing the system's potential as a fundamental component of a scalable quantum network for building on-chip quantum information processing devices.

\begin{figure}
      \epsfig{file=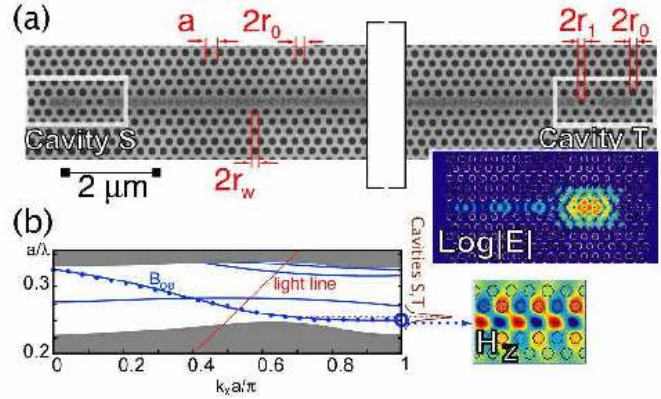, width=4in}
    \caption{Coupled cavities system.  (a) The identical source (S) and target (T) cavities are connected via the 25-$\mu$m waveguide and can be probed and pumped separately. The design parameters are: PC, $a=256$\unit{nm}, $r_0=0.3 a$; outer cavities, $r_1=0.25 a, r_0=0.3 a$; waveguide, $r_w=0.25 a$ for the bounding rows of holes. \textit{Inset:} Electric field pattern. (b) The waveguide dispersion diagram shows four TE modes of different symmetries; the $B_{oe}$ band is used for photon transfer between cavities.  The cavity resonances are positioned so that they intersect the lower tail of the waveguide band's linear dispersion region and confine the mode near $k_x=\pi/a$ (field pattern shown in inset).}
    \label{fig:Fig2}
\end{figure}

The structure consists of two linear 3-hole defect cavities\cite{Noda2003Nature}, butt-coupled and connected via a $25$\unit{$\mu$m}-long closed portion of a waveguide (Fig. \ref{fig:Fig2}(a)).  It was designed by component-wise Finite Difference Time Domain (FDTD) simulations in three dimensions. The waveguide design shown here supports four modes; we picked one of these modes, designated as $B_{oe}$ in Fig. \ref{fig:Fig2}(b), to transfer light between the end-cavities.  This mode offers a wide, relatively flat spectral region of guided modes for coupling and can furthermore be confined in the high-$Q$ end-cavities. Thanks to their near-minimum mode volume $V_{mode} \equiv (\int_V \varepsilon(\vec{r}) |\vec{E}(\vec{r})|^{2} d^{3}\vec{r})/\max(\varepsilon(\vec{r}) |\vec{E}(\vec{r})|^{2})\approx 0.74 (\lambda/n)^{3}$, these end-cavities allow a large SE rate enhancement $\propto Q/V_{mode}$.

The cavity and waveguide field decay rates can be expressed as a sum of vertical, in-plane, and material loss, respectively: $\kappa=\kappa_{\perp}+\kappa_{\|}+\gamma$.  Removed from the waveguide, the `bare' outer cavities radiate predominantly in the vertical direction at rate $\kappa_{\perp}$, as in-plane losses can be suppressed with enough PC confinement layers.  Introducing a waveguide coupled to the cavity creates additional loss $\kappa_{\|}$.  In designing the cavity/waveguide system, we therefore optimized for the ratio of the coupling rate into the waveguide versus other losses, $\kappa_{\|}/(\kappa_{\perp}+\gamma)$, while retaining a high cavity $Q$ value for enhancing the SE rate of QDs coupled to the cavity.

In a waveguide of finite extent, the continuum of modes in the waveguide band breaks up into discrete resonances.  For photon transfer, one of these must be coupled to the outer cavities.  Assuming spectral matching and negligible material losses\cite{Englund05_OptExp}, the field amplitudes in the source and target cavities (S,T) and waveguide (W) evolve according to:
\begin{eqnarray}
\label{eq:rates}
 \nonumber \dot{c_s}(t) &=& -i \kappa_{\|} c_w(t) - \kappa_{\perp} c_s(t) + p(t) \\
\dot{c_t}(t) &=& -i \kappa_{\|} c_w(t) - \kappa_{\perp} c_t(t)  \\
\dot{c_w}(t) &=& -i\kappa_{\|} c_s(t) -i\kappa_{\|} c_t(t) - \kappa_W c_w(t) \nonumber
\end{eqnarray}
Here we assume equal coupling rates for the outer cavities, based on their near-identical SEM images and on their $Q$ values, which fall within a linewidth of each other in the great majority of structures.  The constant $\kappa_W$ denotes the waveguide loss rate (other than loss into the end-cavities), and $p(t)$ represents a dipole driving the source cavity.  It will suffice to analyze this system in steady-state since excitation of the modes, which happens on the order of the exciton lifetime $\tau\sim 100$\unit{ps}, occurs slowly compared to relaxation time of the photonic network, which is of order $\tau_{cav}=\omega/Q \sim 1$\unit{ps} for the cavity and waveguide resonances involved.  Then the amplitude ratio between the S and T fields is easily solved to be $|c_s/c_t|=1+\kappa_{\perp}\kappa_W/\kappa_{\|}^{2}$.

In the present application, where high photon transfer probability and reasonably high output coupling is desired, we optimized for a design with a two-hole separation between cavity and waveguide, giving $Q_{\perp} \equiv 2 \omega/\kappa_{\perp}=23000, Q_{\|}\equiv 2 \omega/\kappa_{\|}=5200$, and $\kappa_{\|}/\kappa_{\perp}=4.4$.  In this design, the cavity resonances were targeted to a linear region of the waveguide dispersion, slightly above the lower waveguide cut-off frequency (see Fig.\ref{fig:Fig2}).  In this way, we ensured coupling that is tolerant to slight fabrication inaccuracies.


The structures were fabricated on a 160\unit{nm} GaAs membrane, grown by molecular-beam epitaxy with a central layer of self-assembled InAs QDs whose photoluminesce peaks at 932\unit{nm} with an inhomogeneous linewidth of $\sim 60$\unit{nm} and a density of 200\unit{QDs$/\mu$m$^{2}$}.  The structures were fabricated by electron beam lithography and reactive ion etching, followed by a wet chemical etch to remove a sacrificial layer underneath the PC membrane\cite{Englund05PRL}. Fabrication inaccuracies decreased $Q$ values to about $1000-5000$.

\begin{figure}
\epsfig{file=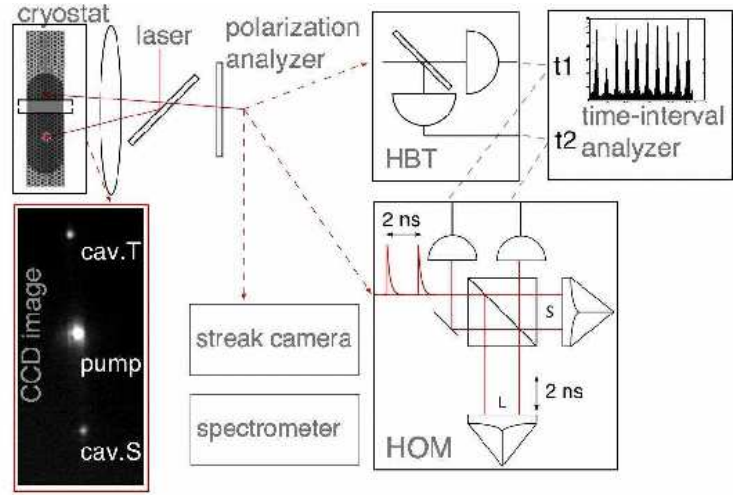, width=4in}
    \caption{ The experimental setup.  The sample, positioned in the He-flow cryostat at 5K, is addressed with a confocal microscope with a stir-able pump beam (spot diameter $\sim 1$\unit{$\mu$m}) and movable probe aperture (selection region diameter 2.9\unit{$\mu$m}).  The microscope directs photoluminescence to the 0.75\unit{m} spectrometer (with liquid-nitrogen cooled Si detector),  0.75\unit{m} streak camera, or Hanbury-Brown-Twiss or Hong-Ou-Mandel setups. We estimate the coupling efficiency from cavity to external optics at $\sim11$\%.}
    \label{fig:Fig3}
\end{figure}
\begin{figure}
 \epsfig{file=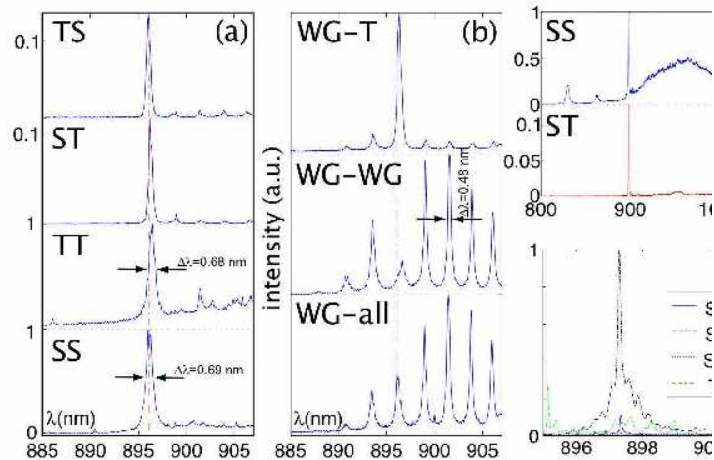, width=4in}
    \caption{Cavity-cavity coupling via a waveguide.  (a) Direct and transferred cavity emissions. Pumping and observing the same cavities show high spectral overlap between the cavities (plots `SS' and `TT').  When cavity S (T) is pumped (above-band at 780\unit{nm}), the transferred emission is observed from cavity T (S), as shown in plot `ST' (`TS').  In this high-power excitation, broad-band emission that includes higher-order transitions maps out the spectrum.  The field intensity ratio between source and target cavities is estimated at 0.12 from the ratios of the heights of curves `SS' and `ST'.  (b) Coupling occurs through spectral overlap of the cavities with a waveguide resonance.  Here the waveguide is pumped and emission collected from the full structure (`WG-all') or spectrally filtered from the waveguide (`WG-WG') or cavity T (`WG-T').  Plot `WG-T' clearly shows that the high spatial selectivity allows collection only from the cavity; plot `WG-WG' indicates drop-filtering into T.  (c) Broad emission from above-band excitation of QDs in cavity S (plot `SS') is filtered into the target cavity (plot `ST').    (d) When the QD exciton at 897.3\unit{nm} in cavity S is pumped (resonantly at 878\unit{nm}, 460\unit{$\mu$W}), the emission is observed from S (`SS') and T (`ST').  The cross-polarized spectrum from S shows nearly complete quenching of QD emission (`SS, $90^{\circ}$').  The line at 897.3\unit{nm} is only observed when S is pumped --- e.g., it is not visible when T is pumped and T observed (`TT').  }
    \label{fig:Fig4}
\end{figure}

We will now focus on a particular system that showed high coupling among cavities, while simultaneously evincing large QD coupling in the source cavity.  Measurements were done with the sample at 5K in a continuous-flow cryostat and probed with the confocal microscope setup connected to the various instruments shown in Fig. \ref{fig:Fig3}.  A movable aperture is located the image plane of the microscope so that the pump and observation regions can be adjusted independently.  The structures were characterized by measuring the combinations of pump/probe regions (waveguide `WG', source cavity `S', or target cavity `T').  Here the broad-band photoluminescence (PL) of high-intensity, above-band pumped QDs was used as an internal illumination source. Fig. \ref{fig:Fig4}(a) shows good spectral match between direct measurements of the source/target cavities (plots `SS' and `TT'), together with the coupled emission (plots `ST' and `TS').   A comparison of the emission intensities from the S and T cavities gives the transfer efficiency $|c_t/c_s|^{2}=0.12$.  To understand how this transmission occurs through the terminated waveguide, we illuminate the waveguide resonances by pumping it near the center.  Photoluminescence that is resonant inside the waveguide, but off-resonant from the cavities, is scattered primarily at the waveguide/cavities interfaces.  On the other hand, PL that is resonant with the cavities is dropped into them and can be spatially separated with the pinhole.  This drop-filtering is shown in \ref{fig:Fig4}(b), where the bottom plot (`WG-all') shows all modes resonant in the coupled system, while the top two plots `WG-WG' and `WG-T' show collection from only the waveguide terminus and the cavity, respectively.  Fitting the measurements of panels (a,b) to the frequency-domain model in Eq.\ref{eq:rates} gives the coupling coefficients $\kappa_{\perp}=455$\unit{GHz}, $\kappa_{W}=322$\unit{GHz}, $\kappa_{\|}=283$\unit{GHz}.  Other instances of a slightly modified cavity/waveguide design, in which the waveguide and cavity were separated by only a single hole, yielded photon transfer ratios as large as $|c_t/c_s|^{2}=0.49$ (see supplemental material), though we did not produce this system in large enough numbers to find high cavity/QD coupling. The cavities/waveguide system strongly isolates transmission to the cavity linewidth, as seen from the transmission spectrum (`ST') (Fig. \ref{fig:Fig4} inset) when S is pumped above the GaAs bandgap.



\begin{figure}
 \epsfig{file=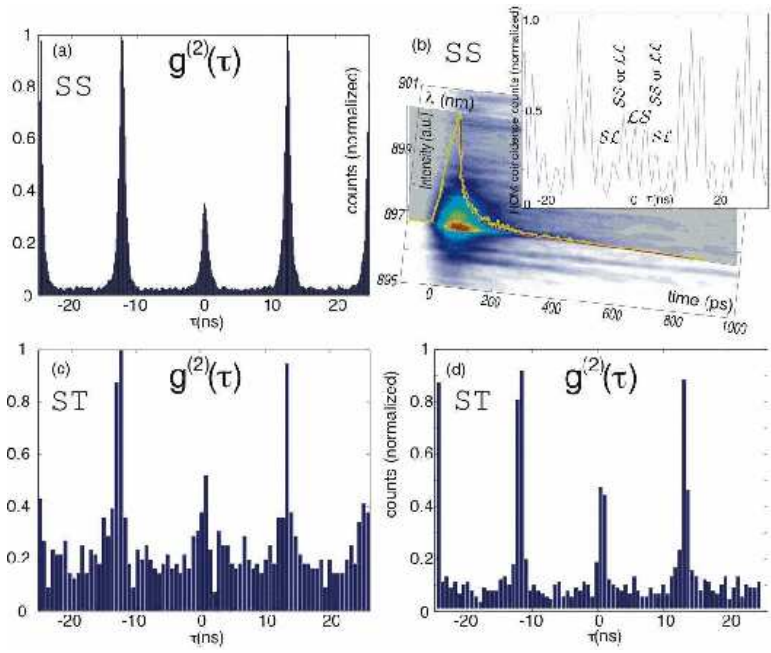, width=4in}
    \caption{Single photon source characterization.  (a) Autocorrelation data when cavity S pumped, cavity S collected. (b) Streak camera data indicate exciton lifetime $\tau=116$\unit{ps}.  The rise-time (jitter-time) is measured at 23\unit{ps} with a lower-resolution grating with higher time response (data not shown).  \textit{Inset:} Two-photon interference experiment.  Pairs of photons, emitted 2.3\unit{ns} apart and repeated every 13\unit{ns}, enter the unbalanced Michelson interferometer with long path $\mathcal{L}$ and short path $\mathcal{S}$ (see Fig. \ref{fig:Fig3}).  Peak $\mathcal{LS}$ corresponds to the first photon going through $\mathcal{L}$ and the second through $\mathcal{S}$, so that they collide on the beamsplitter.  With increasing indistinguishability, these photons experience higher interference, which results in a decreased area of peak $\mathcal{LS}$. The non-zero area of peak $\mathcal{LS}$ results largely from non-zero $g^{(2)}(0)$ of the source.  (c) Autocorrelation data when cavity S pumped, cavity T collected (with grating filter). (d) Cavity S pumped and T collected directly (no grating filter). }
    \label{fig:Fig5}
\end{figure}

We now consider the problem of coupling a QD to the source cavity, which requires a high degree of spatial and spectral matching.  Though it primarily relies on chance, the spectral coupling can be fine-tuned by shifting the QD transitions with changing sample temperature. In Fig. \ref{fig:Fig4}(d), we show a single-excition transition coupled to cavity S at 897\unit{nm}.  The transition is driven resonantly through a higher-order excited QD state with a 878\unit{nm} pump from a Ti-Saph laser.  The SE rate enhancement is measured from the modified emitter lifetime, which a direct streak camera measurement puts at 116\unit{ps} (Fig. \ref{fig:Fig5}(b)).  Compared to the average lifetime of 1.4\unit{ns} for QDs in the bulk semiconductor of this wafer, this corresponds to a Purcell enhancement of $F=12$. The SE coupling factor into the cavity mode is then $\beta=F/(F+F_{PC})\sim0.98$, where $F_{PC}\sim 0.3$ reflects the SE rate suppression into other modes due to the bandgap of the surrounding PC\cite{Englund05PRL}.


We characterized the exciton emission by measurements of the second-order coherence and indistinguishability of consecutive photons.  The second-order coherence $g^{(2)}(t')=\braket{I(t) I(t+t')}/\braket{I(t)}^{2}$ is measured with a Hanbury-Brown and Twiss interferometer, as described earlier \cite{Englund05PRL}.  When the QD in cavity S is pumped resonantly, then photons observed from S shows clear antibunching (Fig. \ref{fig:Fig5}(a)), with $g^{2}(0)=0.35\pm0.01$.  This value is larger than what we observe for resonantly pumped cavity-detuned QDs, where typically $g^{(2)}(0)<0.05$, similar to previous reports\cite{Santori02}.  The main contributor to the larger $g^{(2)}(0)$ for the coupled QD is enhanced background emission from nearby transitions and the wetting layer emission tail, which decays at $\sim 100$\unit{ps} time scales (see Fig. \ref{fig:Fig5}(b)) and is not completely filtered by our grating setup.  The background emission is rather large in this study because of the high QD density of the sample (e.g., four times larger than that in the experiment by \textit{Santori et al.}\cite{Santori02}).  Because of the shortened lifetime of the cavity-coupled QD exciton, the coherence time of emitted photons becomes dominated by radiative effects and results in high photon indistinguishability\cite{Kiraz04}.  We measured the indistinguishability using the Hong-Ou-Mandel (HOM) type setup sketched in Fig. \ref{fig:Fig3}, similar to a recent experiment on PC-cavity coupled QDs\cite{Abram2005APL}.  Following the analysis of \textit{Santori et al.}\cite{Santori02}, the data (inset Fig.\ref{fig:Fig5}(b)) indicate a mean wavefunction overlap of $I=0.67\pm0.18$, where we adjusted for the imperfect visibility (88\%) of our setup and subtracted dark counts in the calculation. Even with higher SE rate enhancement, we expect that $I \lesssim 0.80$\ for resonantly excited QDs\cite{Vuckovic2005PRE} because of the finite relaxation time, measured here at $ 23$\unit{ps} by the streak camera.  

We will now consider the transfer of single photons to the target cavity T.  The single photon transfer is described by Eqs. \ref{eq:rates}, where cavity S is now pumped by the QD single exciton with state $e(t) \ket{e}+ g(t)\ket{g}$:
\begin{eqnarray}
p(t)&=&-i g_0 e(t) \\ \nonumber
\dot{e}(t)&=&-\fr{\Gamma}{2} e(t)-i g_0 g(t)
\end{eqnarray}
Here, $g_0$ is the QD-field coupling strength and $\Gamma$ the QD spontaneous emission rate.  In the present situation, where coupling rates greatly exceed the exciton decay rate, the previous steady-state results apply.  Thus, the signal from cavity T mirrors the SE of the single exciton coupled to cavity S.  Experimentally, we verified photon transfer from S to T by spectral measurements as in Fig. \ref{fig:Fig4}(d): the exciton line is observed from T only if S is pumped.  It is not visible if the waveguide or cavity T itself are pumped, indicating that this line originates from the QD coupled to cavity S and that a fraction of the emission is transferred to T.  This emission has the same polarization and temperature-tuned wavelength dependence as emission from S.  Photon autocorrelation measurements on the signal from T indicate the antibunching characteristic of a single emitter when S is pumped (Fig. \ref{fig:Fig5}(c)).  The signal-to-noise ratio is rather low because autocorrelation count rates are $(|c_t/c_s|^{2})^{2} \sim 0.014$ times lower than for collection from S.  Nevertheless, the observed antibunching does appear higher, in large part because the background emission from cavity S is additionally filtered in the transfer to T, as shown in Fig.\ref{fig:Fig4}(b).  Indeed, this filtering through the waveguide/cavity system suffices to bypass the spectrometer in the HBT setup (a 10-nm bandpass filter was used to eliminate room lights).  Slightly more background is permitted, but the count rate is about three times higher while antibunching, $g^{2}(0)=0.50\pm0.11$, is still clearly evident (Fig. \ref{fig:Fig5}(d)).   The largest contribution to $g^{2}(0)$ comes from imperfectly filtered photoluminesce near the QD distribution peak seen in Fig. \ref{fig:Fig4}(c).  This transmission appears to occur through the top of the dielectric band near $k=0.7 \pi/a$, and could easily be eliminated in the future by increasing the waveguide frequency with slightly larger bounding holes around the PC cavity.  This on-chip filtering will be essential in future QIP applications and should also find uses in optical communications as a set of cascaded drop filters.  Table 1\ref{tab:Tab1} summarizes the relevant parameters of this system.



\begin{table}
\label{tab:Tab1}
\begin{center}
\begin{tabular}{l|c|c|c|c|c|c}
  &$g_0$ (GHz) & $\kappa$ (GHz) & $\kappa_{\|}/\kappa_{\perp}$ & $\Gamma $(GHz) & $I$\\
 \hline
system 1 & 50 & 800 & 0.62 & 10  & 0.67 \\
best observed & & 210 & 1.9 &  & \\  
theoretical & 230 & 46 & 4.4 & $\sim180^{\dagger}$ &  \\
\end{tabular}
\caption{Paramaters of the structure described here (`system 1'), best observed values from other structures on the same sample (if different), and theoretical predictions of the design.  $^{\dagger}$ \textit{Limited by the onset of strong coupling.} }
\end{center}

\end{table}%

In conclusion, we have demonstrated a basic building block of a quantum network consisting of a quantum dot with large coupling to a high-$Q/V$ cavity, which in turn is coupled to a target cavity via a waveguide.  This system functions as an efficient on-demand source of single photons with mean wavepacket overlap of $\sim 67$\%, SE coupling efficiency $\beta\sim0.98$ into the cavity mode, and high out-coupling efficiency into the waveguide.  These single photons from cavity S are channeled to the target cavity, as confirmed by localized spectroscopic measurements.  We measured a high photon transfer with a field intensity ratio of $0.12$ for this system.  In other structures we measured field ratios up to 49\%, though without coupled QDs.  These efficiencies greatly exceed what is possible in off-chip transfer and demonstrate the great potential of this system as a building block of future on-chip quantum information processing systems.

 Financial support was provided by the MURI Center for photonic quantum information systems (ARO/DTO Program DAAD19-03-1-0199) and NSF grants ECS-0424080 and ECS-0421483. Dirk Englund was also supported by the NDSEG fellowship.  Dr.Zhang is supported by JST.  We thank David Fattal and Edo Waks for helpful discussions.

\bibliographystyle{unsrt}
\bibliography{0603_CcSPS_Arxiv_V3}
\clearpage

\begin{figure}
 \epsfig{file=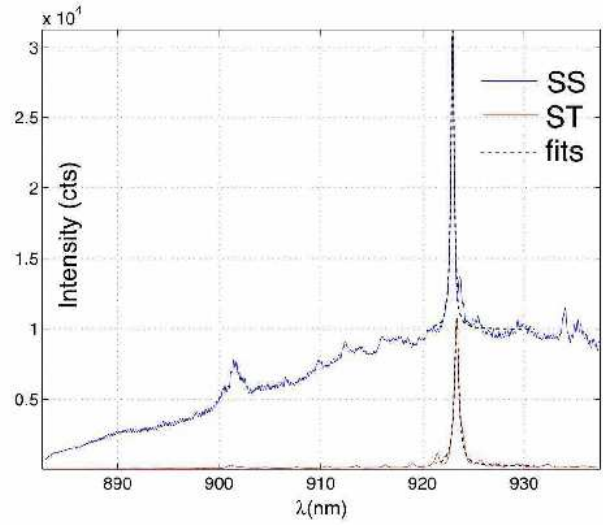, width=4in}
    \caption{\textit{Supplemental Material:} The slightly modified waveguide-coupling design with a single separating hole between cavities and waveguide yields higher coupling -- in this case, the S/T field intensity ratio is estimated at $0.49$. No single QD was coupled to S in this structure.  }
    \label{fig:Fig1_suppl}
\end{figure}

\end{document}